\documentclass[doublecol]{epl2} 

\usepackage{fancyhdr}
\title{First order transition in trigonal structure ${\textbf{Ca}}{\textbf{Mn}}_{2}{\textbf{P}}_{2}$ }
\shorttitle{Title} 

\author{Y. J. Li\inst{1,2} \and F. Jin\inst{1,2} \and Z. Y. Mi\inst{1,2} \and J. Guo\inst{1,2} \and W. Wu\inst{1,2} \and Z. H. Yu\inst{3} \and D. S. Wu\inst{1,2} \and S. H. Na\inst{1,2} \and C. Mu\inst{1,2} \and X. B. Zhou\inst{1} \and Z. Li\inst{1,2} \and K. Liu\inst{4} \and L. L. Sun\inst{1,2,5} \and Q. M. Zhang\inst{6,1} \and T. Xiang\inst{1,2,7} \and G. Li\inst{1,2,5} \and J. L. Luo\inst{1,2,5}\thanks{E-mail: \email{jlluo@iphy.ac.cn (corresponding author)}}}
\shortauthor{Y. J. Li \etal}

\institute{
  \inst{1} Beijing National Laboratory for Condensed Matter Physics and Institute of Physics, Chinese Academy of Sciences, Beijing 100190, China\\
  \inst{2} School of Physical Sciences, University of Chinese Academy of Sciences, Beijing 100190, China\\
  \inst{3} School of Physical Science and Technology, ShanghaiTech University, Shanghai 201210, China\\
  \inst{4} Department of Physics and Beijing Key Laboratory of Opto-electronic Functional Materials \& Micro-nano Devices, Renmin University of China, Beijing 100872, China\\
  \inst{5} Songshan Lake Materials Laboratory, Dongguan, Guangdong 523808, China\\
  \inst{6} School of Physical Science and Technology, Lanzhou University, Lanzhou 730000, China\\
  \inst{7} Kavli Institute for Theoretical Sciences, Beijing 100190, China
}

\abstract{
We report structural and physical properties of the single crystalline ${\mathrm{Ca}}{\mathrm{Mn}}_{2}{\mathrm{P}}_{2}$. The X-ray diffraction(XRD) results show that ${\mathrm{Ca}}{\mathrm{Mn}}_{2}{\mathrm{P}}_{2}$ adopts the trigonal ${\mathrm{Ca}}{\mathrm{Al}}_{2}{\mathrm{Si}}_{2}$-type structure. Temperature dependent electrical resistivity $\rho(T)$ measurements indicate an insulating ground state for ${\mathrm{Ca}}{\mathrm{Mn}}_{2}{\mathrm{P}}_{2}$ with activation energies of 40 meV and 0.64 meV for two distinct regions, respectively. Magnetization measurements show no apparent magnetic phase transition under 400 K. Different from other ${\mathrm{A}}{\mathrm{Mn}}_{2}{\mathrm{Pn}}_{2}$ (A = Ca, Sr, and Ba, and Pn = P, As, and Sb) compounds with the same structure, heat capacity $C_{\mathrm{p}}(T)$ and $\rho(T)$ reveal that ${\mathrm{Ca}}{\mathrm{Mn}}_{2}{\mathrm{P}}_{2}$ has a first-order transition at $T$ = 69.5 K and the transition temperature shifts to high temperature upon increasing pressure. The emergence of plenty of new Raman modes below the transition, clearly suggests a change in symmetry accompanying the transition. The combination of the structural, transport, thermal and magnetic measurements, points to an unusual origin of the transition.}

\begin{document}

\maketitle

\section{Introduction}
Exploring new superconductors in the 3d transition-metal-based compounds is one of the most active research topics in modern condensed matter physics. Owning to the interplay among charge, spin, and orbital degrees of freedom, various ground states could be stabilized in these materials.  Superconductivity has been observed in the majority of 3d transition-metal-based compounds, however it is still rare for Cr- and Mn-based ones. Up to now, a few Cr-based superconductors has been reported at either ambient or elevated pressure, such as CrAs \cite{wu2014superconductivity}, quasi one-dimensional ${\mathrm{A}}_{2}{\mathrm{Cr}}_{3}{\mathrm{As}}_{3}$, ${\mathrm{A}}{\mathrm{Cr}}_{3}{\mathrm{As}}_{3}$ (A=Na, K, Rb, Cs) and ${\mathrm{Ln}}_{3}{\mathrm{Cr}}_{10-x}{\mathrm{N}}_{11}$ (Ln=La, Pr)\cite{bao2015superconductivity,tang2015unconventional,Mu2017PhysRevB,Shao2018EurophysLett,Wu2019NationalScienceReview,Li2019EurophysLett,Nigro2019EurophysLett}. Meanwhile, to our knowledge, there is only one Mn-based superconductor, MnP \cite{Cheng2015PRL}. It is of great interest to explore other possible Mn-based superconductors\cite{Na2020CPL}.

The body-centered tetragonal ternary compounds ${\mathrm{A}}{\mathrm{Fe}}_{2}{\mathrm{As}}_{2}$ (A=Ca, Sr, Ba, Eu) are metallic and show nearly contiguous antiferromagnetic spin-density wave(SDW) and structure transitions \cite{Rotter2008PRB,Chen2008PRB,Krellner2008PRB,Ni2008PRB,Ronning2008JPCM1,Goldman2008PRB,Tegel2008JPCM,Ren2008PRB}. The suppression of these transitions by external pressure or chemical doping leads to superconductivity with bulk transition temperature, $T_{c}$ up to 45 K \cite{Chen2008CPL,Rotter2008PRL,Kimber2009NM,Wu2013PNAS}. The iron atoms can be fully replaced by other transition metal atoms, such as Cr, Mn, Co, Ni, and Cu \cite{Sangeetha2016PRB1,Brock1994JSSC,Sefat2009PRB,Li2014JPCM,Saparov2012JSSC}. Among them, ${\mathrm{Sr}}{\mathrm{Ni}}_{2}{\mathrm{As}}_{2}$ and ${\mathrm{Ba}}{\mathrm{Ni}}_{2}{\mathrm{As}}_{2}$ were found to be superconductors with $T_{c}$ = 0.62 and 0.7 K, respectively \cite{Bauer2008PRB,Ronning2008JPCM2}.

The Mn pnictides ${\mathrm{A}}{\mathrm{Mn}}_{2}{\mathrm{Pn}}_{2}$ (A=alkaline earth, Pn=P, As, Sb) have been known for decade. Different from ${\mathrm{Ba}}{\mathrm{Mn}}_{2}{\mathrm{Pn}}_{2}$ crystallizing in the tetragonal ${\mathrm{Th}}{\mathrm{Cr}}_{2}{\mathrm{Si}}_{2}$ structure type ($I4/mmm$) similar to ${\mathrm{A}}{\mathrm{Fe}}_{2}{\mathrm{As}}_{2}$, the compounds ${\mathrm{Ca}}{\mathrm{Mn}}_{2}{\mathrm{As}}_{2}$, ${\mathrm{Ca}}{\mathrm{Mn}}_{2}{\mathrm{Sb}}_{2}$, and ${\mathrm{Sr}}{\mathrm{Mn}}_{2}{\mathrm{Pn}}_{2}$ crystalize in the trigonal ${\mathrm{Ca}}{\mathrm{Al}}_{2}{\mathrm{Si}}_{2}$ structure ($P\bar{3}m1$). Previous reports suggest that they are all insulating at low temperature and have antiferromagnetic (AFM) order \cite{Brock1994JSSC,Wang2011JPCS,Sangeetha2016PRB,Das2017JPCM,Simonson2012PRB,Gibson2015PRB,Singh2019PRB,Saparov2013JSSC}. However, for ${\mathrm{Ca}}{\mathrm{Mn}}_{2}{\mathrm{P}}_{2}$, apart from the structural data, no detailed physical properties have been reported so far, which gives us motivation to carry out current study.

In this work, we report the growth, crystal structure, electrical resistivity $\rho(T)$ in the $ab$ plane at ambient and elevated pressure, Raman spectra, magnetic susceptibility $\chi(\mathrm{T})$, and heat capacity $C_{\mathrm{p}}(T)$ measurements of ${\mathrm{Ca}}{\mathrm{Mn}}_{2}{\mathrm{P}}_{2}$ single crystals. It is found that ${\mathrm{Ca}}{\mathrm{Mn}}_{2}{\mathrm{P}}_{2}$ has a structural type first order transition at 69.5 K, at the same time there is no long range magnetic order developed below 400 K, thus it is different from other isostructural Mn- based pnictides and calls for further study.

\section{Experiments}
Single crystal of ${\mathrm{Ca}}{\mathrm{Mn}}_{2}{\mathrm{P}}_{2}$ was grown using Sn flux. High purity elements Ca (99\%), Mn (99.99\%), P (99.999\%), and Sn (99.9999\%) were weighed in the molar ratio Ca:Mn:P:Sn =1.05:2:2:20 and placed in an alumina crucible. The crucible was then sealed into a quartz tube under an Ar pressure of 1/4 ATM. After preheating the mixture at 300${ }^{\circ} \mathrm{C}$ for 24h, the assembly was heated to 1100${ }^{\circ} \mathrm{C}$, dwelled at the temperature for 24 hours for homogenization, and subsequently cooled down to 700${ }^{\circ} \mathrm{C}$ at the rate of 2${ }^{\circ} \mathrm{C}$/h. Finally the tube was taken out of the furnace quickly to separate the tin flux by centrifuging. Shiny, plate-like single crystals in hexagon shape with maximum dimensions 3$\times$ 1.5$\times$ 1$\mathrm{mm}^{3}$ were obtained.

Powder XRD experiments were conducted by a PAN-analytical X-ray diffractometer with Cu $\mathrm{K}_{\alpha}$ radiation, the angle 2$\theta$  range was from 10$^{\circ}$ to 80$^{\circ}$. The element analysis was performed by using an energy dispersive X-ray spectroscopy (EDX) equipped on a Hitachi S-4800 scanning electron microscope (SEM). Single crystal XRD structural analysis of ${\mathrm{Ca}}{\mathrm{Mn}}_{2}{\mathrm{P}}_{2}$ was performed at room temperature and 40 K in a Bruker D8 Venture diffractometer using Mo $\mathrm{K}_{\alpha}$ radiation, $\lambda$ =0.71073 \AA. Structure refinement was performed by the program SHELXL-2014/7 \cite{Sheldrick2015Crystallogr} embedded in the program suite APEX3.

The electrical resistivity and specific heat were measured in a Quantum Design physical property measurement system (PPMS-9) by the standard four-probe method with electrodes along the $ab$ plane and relaxation method, respectively. High-pressure resistance measurements were performed in a diamond-anvil cell (DAC), in which diamond anvils with 300 $\mu m$ flats and a nonmagnetic rhenium gasket with 100 $\mu m$ diameter hole were employed. NaCl powder was used as the pressure medium to provide a quasi-hydrostatic pressure environment. The temperature dependence of DC magnetic susceptibility was measured in a Quantum Design magnetic property measurement system (MPMS).

The polarized Raman spectra were collected with a Jobin Yvon LabRam HR Evolution spectrometer equipped with a volume Bragg grating low-wave-number suite, a liquid-nitrogen-cooled back-illuminated charge-coupled device detector, and a 532-nm laser. The laser was focused into a spot of $\sim$ 5 $\mu m$ in diameter on the sample surface ($ab$ plane), with a power $<$100 $\mu$W, to avoid overheating.

\section{Results And Discussions}
\begin{figure}
\onefigure[scale=0.33]{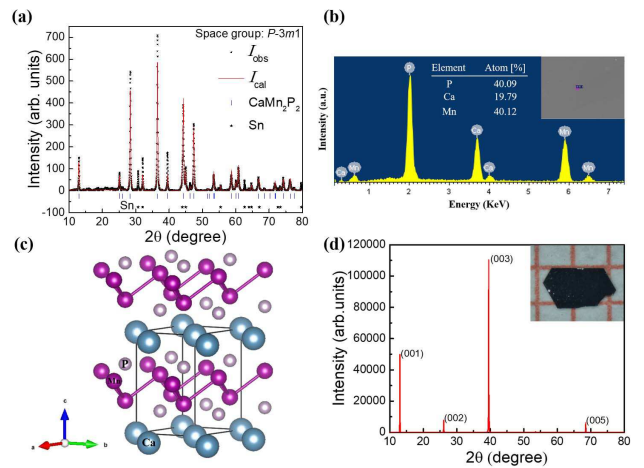}
\caption{Structural and elemental analysis of ${\mathrm{Ca}}{\mathrm{Mn}}_{2}{\mathrm{P}}_{2}$. (a) Powder XRD spectrum of crushed crystals at room temperature with indexing. (b) EDX spectrum. (c) An illustration of the crystal structure. (d) Single crystal XRD spectrum at room temperature, inset shows a photo of grown crystal on milimeter scale. }
\label{fig:structure}
\end{figure}

Powder XRD data on crushed ${\mathrm{Ca}}{\mathrm{Mn}}_{2}{\mathrm{P}}_{2}$ single crystals at room temperature is shown in fig.~\ref{fig:structure}(a)
. The reflections can be well indexed by the trigonal unit cell. However, weak peaks from adventitious Sn flux are also visible and marked by stars. The reason is that due to a pile of single crystals were used to prepare the sample for powder XRD, the Sn residual on the surface of ${\mathrm{Ca}}{\mathrm{Mn}}_{2}{\mathrm{P}}_{2}$ single crystals were not completely removed and entered XRD spectrum. After further cleaning, EDX analysis was performed to check the chemical composition of the grown single crystals. The typical EDX spectrum of an individual crystal is shown in fig.~\ref{fig:structure}(b). Only three elements, Ca, Mn and P are detected. The average ratio of the elements at different locations in the crystals is 19.79:40.12:40.09, which is close to the 1:2:2 stoichiometry of the compound. The Sn flux and any other possible impurity were not detected in present analysis within instrumental sensitivity.

A tiny piece of single crystal was used for the structure refinement at room temperature. R1 and wR2 are residual factors. The final anisotropic full-matrix least-squares refinement on $F^{2}$ (F is structural factor) with 10 variables converges at R1 for the observed data and wR2 for all data\cite{Sheldrick2015Crystallogr}. The result confirms that the crystal has the trigonal ${\mathrm{Ca}}{\mathrm{Al}}_{2}{\mathrm{Si}}_{2}$-type structure with space group $P\bar{3}m1$, and the refined lattice constants are a=4.0973(1) \AA, c=6.8473(3) \AA, respectively. The refinement parameters are tabulated in table~\ref{tab:table1},
\begin{table}
\small
\caption{Refined crystallographic parameters of ${\mathrm{Ca}}{\mathrm{Mn}}_{2}{\mathrm{P}}_{2}$ from single crystal X-ray structure analysis at 293K and 40K.}
\label{tab:table1}
\begin{center}
\scalebox{0.89}{
\begin{tabular}{ccc}
\hline \hline
\textrm{Refined parameter}& \textrm{293K}&\textrm{40K}\\
\hline
Wavelength & 0.71073\AA & 0.71073\AA\\
Structure & ${\mathrm{Ca}}{\mathrm{Al}}_{2}{\mathrm{Si}}_{2}$-type & ${\mathrm{Ca}}{\mathrm{Al}}_{2}{\mathrm{Si}}_{2}$-type \\
                           &  trigonal &  trigonal \\
Space group & $P\bar{3}m1$ (No.164) & $P\bar{3}m1$ (No.164)\\
Lattice parameters & a=4.0973(1) \AA  & a=4.0887(1) \AA \\
                   & b=4.0973(1) \AA  & b=4.0887(1) \AA \\
                   & c=6.8473(3) \AA  & c=6.8207(3) \AA \\
                   & $\alpha$=90$^{\circ}$   & $\alpha$=90$^{\circ}$ \\
                   & $\beta$=90$^{\circ}$    & $\beta$=90$^{\circ}$ \\
                   & $\gamma$=120$^{\circ}$  & $\gamma$=120$^{\circ}$ \\
Absorption coefficient & 8.187 mm$^{-1}$ & 8.253 mm$^{-1}$\\
R1 & 2.37\% & 1.4\%\\
wR2 & 5.7\% & 3.3\%\\
The goodness-of fit & 1.412 & 1.167\\
Ca1 & (0,0,0)               & (0,0,0)\\
Mn1 & (0.667,0.333,0.625) & (0.667,0.333,0.624)\\
P1  & (0.667,0.333,0.261) & (0.667,0.333,0.262)\\
\hline
\end{tabular}
}
\end{center}
\end{table}
which are in accordance to the early report \cite{Mewis1978Naturforsch}.

The crystal structure of trigonal ${\mathrm{Ca}}{\mathrm{Mn}}_{2}{\mathrm{P}}_{2}$ is shown in fig.~\ref{fig:structure}(c). Similar to other 122-type Mn pnictide compounds, it consists of corrugated honeycomb [${\mathrm{Mn}}_{2}{\mathrm{P}}_{2}$]$^{-2}$ layers that are stacked along the $c$ axis and separated by Ca$^{+2}$ cations. Alternatively, the Mn sublattice can be viewed as triangular double layers of Mn stacked along the $c$ axis and separated by Ca atoms. The XRD results of ${\mathrm{Ca}}{\mathrm{Mn}}_{2}{\mathrm{P}}_{2}$ single crystal are shown in fig.~\ref{fig:structure}(d). Only four main peaks of (001), (002), (003) and (005) are clearly visible, whose positions are in good agreement with those in powder XRD data, indicating that the crystal is grown along the c-axis direction.

\begin{figure}
\onefigure[scale=0.6]{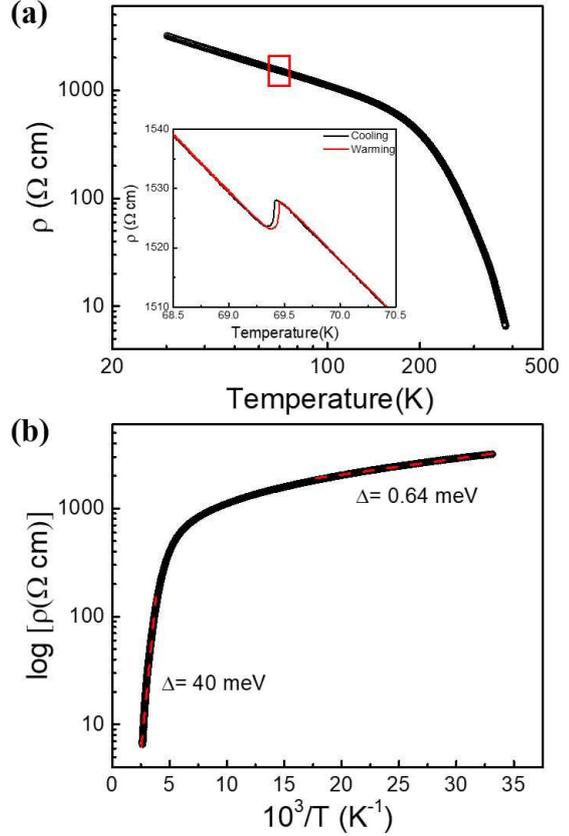}
\caption{(a) Temperature dependence of in-plane electrical resistivity $\rho(T)$ of ${\mathrm{Ca}}{\mathrm{Mn}}_{2}{\mathrm{P}}_{2}$ from 30 K to 400 K. The inset shows the hysteresis resistivity for ${\mathrm{Ca}}{\mathrm{Mn}}_{2}{\mathrm{P}}_{2}$ around 69.5 K. (b) The plot of $\log _{10} \rho$ versus 1000/T. The red dash curves through the data are linear fittings(See text). }
\label{fig:RT}
\end{figure}

Figure~\ref{fig:RT}(a)
gives the $ab$ plane $\rho(T)$ from 30 to 400 K. It is clear that ${\mathrm{Ca}}{\mathrm{Mn}}_{2}{\mathrm{P}}_{2}$ has an insulating ground state, on the contrary to the prediction by a semi-local DFT band structure calculation \cite{Jain2013APL}, in which neither magnetic exchange nor on-site Coulomb repulsion were considered. Whether the insulating gap is due to electron correlation as in a Mott insulator requires further study. The plot of $\log _{10} \rho$ versus 1000/T is shown in fig.~\ref{fig:RT} (b). The $\rho(T)$ data over confined temperature interval is fitted by the expression
\begin{equation}
\label{eq:one}
\log _{10} \rho=\mathrm{A}+2.303 \Delta / \mathrm{k_BT},
\end{equation}
where A is a constant, $k_B$ is Boltzmann$^,$s constant, and $\Delta$ is the activation energy, respectively. The data between 250 K and 400 K and between 30 K and 60 K are nearly linear in $T$ and were separately fitted by Eq.~\ref{eq:one} , shown as the dash curves. The high temperature range fit yields an activation energy $\Delta_1$=40 meV, which is of the same order as previously found for ${\mathrm{Ca}}{\mathrm{Mn}}_{2}{\mathrm{As}}_{2}$ (61 meV) and ${\mathrm{Sr}}{\mathrm{Mn}}_{2}{\mathrm{P}}_{2}$ (12.9 meV)\cite{Brock1994JSSC,Das2017JPCM}, and is the intrinsic activation energy of ${\mathrm{Ca}}{\mathrm{Mn}}_{2}{\mathrm{P}}_{2}$. The low temperature range fit gives a $\Delta_2$= 0.64 meV, which should be extrinsic and originated from the energy gap between donor (acceptor) energy levels and the conduction (valence) band \cite{Pearson1949PR}.

As a feature that has not been observed on other 122-type Mn pnictides, there is a drop of resistivity at $T$ = 69.5 K, which is enlarged as the inset of fig.~\ref{fig:RT}(a), the thermal hysteresis is confirmed by repeated measurements with different cooling/warming speeds, suggesting it corresponds to a first-order transition.

\begin{figure}
\onefigure[scale=0.32]{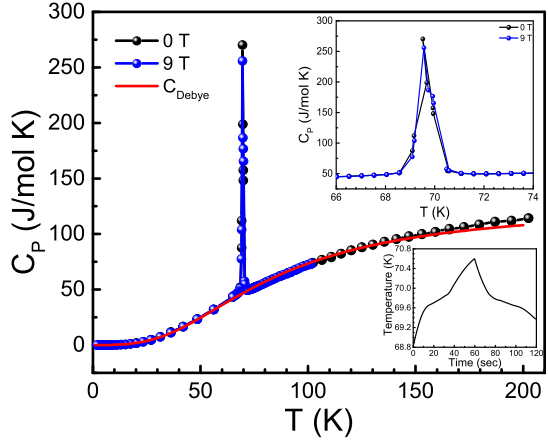}
\caption{Heat capacity $C_{\mathrm{p}}(T)$ of ${\mathrm{Ca}}{\mathrm{Mn}}_{2}{\mathrm{P}}_{2}$ at both 0 T and 9 T, combined with a Debye lattice contribution simulation. The upper inset shows the enlarged region around the first order transition. The lower inset shows a raw temperature relaxation curve demonstrating the existence of latent heat. }
\label{fig:HC}
\end{figure}

The existence of a first-order transition at 69.5 K is further revealed by heat capacity measurement. In the conventional relaxation method, the temperature versus time curves were recorded as the raw data. Shown as the lower inset of fig.~\ref{fig:HC}
 , on a typical fitting curve with the starting temperature is slightly lower than the transition temperature, latent heat is clearly observed on both the raise and the fall processes. Therefore around that temperature a slope analysis is employed to extract $C_p$ value, resulting in the sharp peak in the $C_p$ versus $T$ curve of fig.~\ref{fig:HC}. By applying 9 T magnetic field perpendicular to the crystalline $ab$ plane, neither the position nor the amplitude of the peak is changing, suggesting that the transition is not affected by magnetic fields up to 9 T. Considering the non-metallic ground state previously found from the $\rho(T)$ behavior and the absence of magnetic contribution, the lattice heat capacity alone could account for the total heat capacity at low temperature. We model the $C_{\mathrm{p}}(T)$ data from 2 K to 60 K by
\begin{equation}\label{eq:two}
 C_{p}=C_{lattice}=nC_{Debye}
\end{equation}
in which the Debye lattice heat capacity per mole is
\begin{equation}\label{eq:three}
  C_{D e b y e}=9 R\left(\frac{T}{\Theta_{D}}\right)^{3} \int_{0}^{\Theta_{D} / T} \frac{x^{4} e^{x}}{\left(e^{x}-1\right)^{2}} d x
\end{equation}
where R=8.314 J/mol K is the molar gas constant, and n is the number of atoms per formula unit (n=5 for ${\mathrm{Ca}}{\mathrm{Mn}}_{2}{\mathrm{P}}_{2}$) \cite{Kittel2005book}. The fitting yields Debye temperature $\Theta_{D}$ = 345(1) K and the calculated lattice heat capacity extended up to 200 K is shown as a solid red line superimposed to experimental data in fig.~\ref{fig:HC}.

\begin{figure}
\onefigure[scale=0.9]{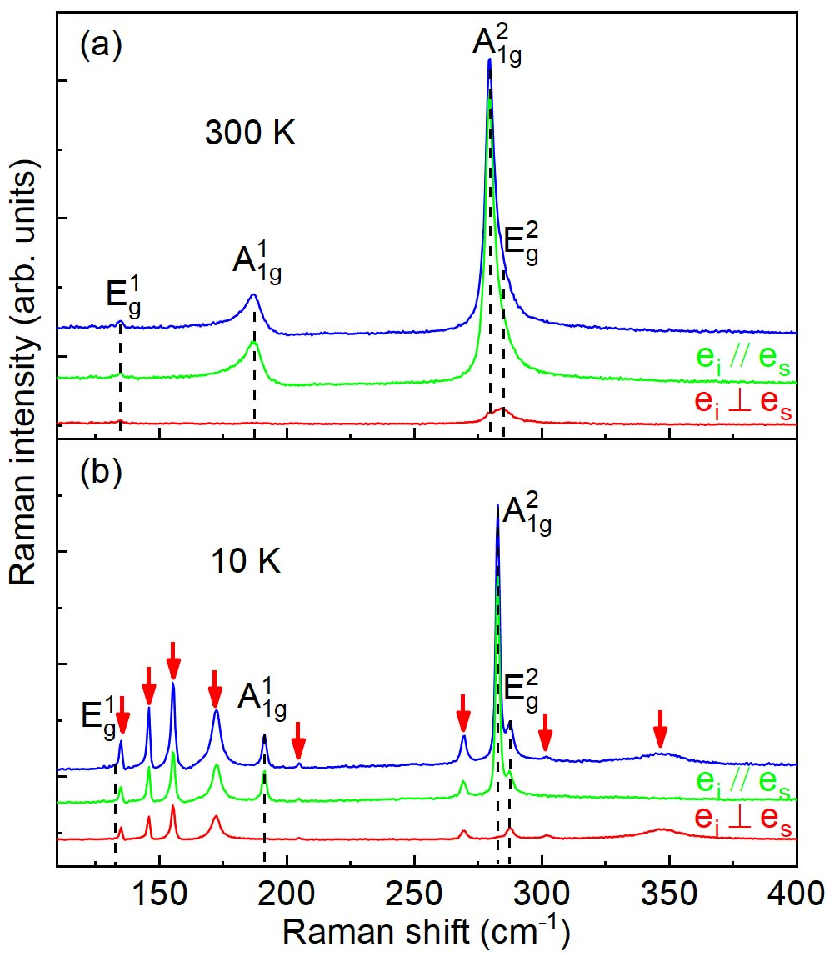}
\caption{Raman spectra of ${\mathrm{Ca}}{\mathrm{Mn}}_{2}{\mathrm{P}}_{2}$ at (a) 300 K (b) 10 K with different polarization. }
\label{fig:Roman}
\end{figure}

In order to better understand the first-order transition of ${\mathrm{Ca}}{\mathrm{Mn}}_{2}{\mathrm{P}}_{2}$, Raman measurements at both 300 K and 10 K are performed on a well characterized crystal, as shown in fig.~\ref{fig:Roman}.
 Four Raman peaks are observed at 300 K, all of which could be assigned according to the symmetry of the crystal structure. However, another 8 new Raman peaks (marked with red arrows) appeared at low temperature. Meanwhile all those high temperature peaks remain present with tiny shift in energy, thus by decreasing temperature through the 69.5 K first order transition, the lattice symmetry is not fully altered, but possibly formed super-structure.

Single crystal XRD structural analysis and powder XRD at temperatures below the first-order transition were also carried out. The refined crystal parameters obtained at 40 K are listed in table~\ref{tab:table1}. From 293 K to 40 K, the lattice constants shrink 0.21\% and 0.39\% for the a- and c-axis, respectively. Meanwhile the lattice space group remains the same. We did not observe any distinguishable peak-splitting nor new peaks at 40 K on the XRD spectrum. Considering the instrumental sensitivity, those results provide further evidence that ${\mathrm{Ca}}{\mathrm{Mn}}_{2}{\mathrm{P}}_{2}$ forms super-structure below 69.5 K.

\begin{figure}
\onefigure[scale=0.29]{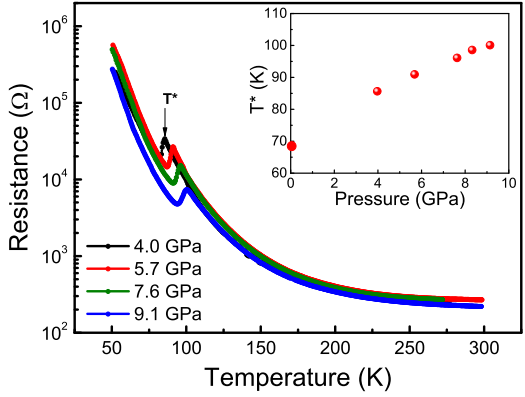}
\caption{Temperature dependence of electrical resistances at different pressures for  ${\mathrm{Ca}}{\mathrm{Mn}}_{2}{\mathrm{P}}_{2}$. The inset shows the pressure dependence of the first-order transition temperature. }
\label{fig:HP}
\end{figure}

Temperature-dependent resistance measurements on single crystals of ${\mathrm{Ca}}{\mathrm{Mn}}_{2}{\mathrm{P}}_{2}$ were performed in a diamond-anvil cell (DAC) with the aim to check whether the compound could be metallized under pressure. However, as shown in fig.~\ref{fig:HP}
, for pressures up to 9.1 GPa, the compound remains an insulating state, and there is no abrupt change in the overall $\rho(T)$ behavior. Meanwhile, the first order transition is robust. Denoting the resistively drop temperature in $\rho(T)$ , as the transition temperature $T^*$, we find that $T^*$ increases almost linearly in the pressure range investigated. It is in contrast to the suppression of AFM transition by external pressure in the metallic 122 iron-arsenides \cite{Kimber2009NM,Wu2013PNAS}. Given that the first order transition is of structural origin, why the low temperature phase becomes more favorable with increasing pressure remains unclear.

\begin{figure}
\onefigure[scale=0.32]{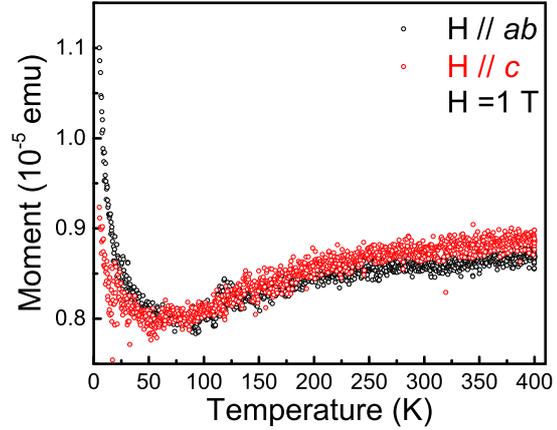}
\caption{Isothermal magnetization of ${\mathrm{Ca}}{\mathrm{Mn}}_{2}{\mathrm{P}}_{2}$ at H=1 T with magnetic field applied along both the $ab$ plane and $c$ axis. }
\label{fig:MT}
\end{figure}

Figure~\ref{fig:MT}
 shows the temperature dependent magnetization of single crystalline ${\mathrm{Ca}}{\mathrm{Mn}}_{2}{\mathrm{P}}_{2}$ measured in a magnetic field $H$=1 T applied in both the $ab$ plane ($H//ab$) and along the $c$-axis($H//c$). Due to the small mass of tested crystal, the data is noisy and not normalized to molar unit. However the overall behavior is clear. The upturn below 30 K is regarded as a contribution from paramagnetic impurities. Other than that, there is no apparent feature at the first order transition temperature, neither any sign of an AFM transition below 400 K. In addition, the anisotropy in magnetization is small. Previous studies on other 122-type Mn pnictides suggested that due to the competition among different exchange interaction energies of Mn sites, these compounds tended to have AFM correlations or develop a long range AFM order. Here for ${\mathrm{Ca}}{\mathrm{Mn}}_{2}{\mathrm{P}}_{2}$, the magnetization increases with increasing temperature, even if a long range AFM order developed at temperature higher than 400 K, the small anisotropy with field applied along two perpendicular directions could rule out a simple AFM type moments alignment along crystal principle axis. Other possible scenarios would include wide fluctuation window of AFM in which the long range magnetic order has not been established or frustration of Mn moment on the hexagonal lattice.

\section{Conclusions}
We have successfully synthesized high quality ${\mathrm{Ca}}{\mathrm{Mn}}_{2}{\mathrm{P}}_{2}$ single crystals by the Sn flux method. The compound is an insulator and crystalizes in the trigonal ${\mathrm{Ca}}{\mathrm{Al}}_{2}{\mathrm{Si}}_{2}$-type structure with lattice parameters $a$=4.0973(1) \AA, and $c$=6.8473(3) \AA \ at room temperature. Heat capacity, Raman spectra and electrical resistivity $\rho(T)$ measurements reveal that ${\mathrm{Ca}}{\mathrm{Mn}}_{2}{\mathrm{P}}_{2}$ undergoes a structural phase transition at $T$=69.5 K and the transition temperature increase with increasing pressure. ${\mathrm{Ca}}{\mathrm{Mn}}_{2}{\mathrm{P}}_{2}$ is unique in that it is the first compound in the hexagonal 122-type Mn pnictides which has a structural transition at low temperature other than apparent magnetic correlation.

\acknowledgments
We are grateful to Youting Song (Institute of Physics) for his help in the single crystal X-ray diffraction experiments. This work was supported by the National Key Research and Development of China (Grant Nos. 2017YFA0302901, and 2017YFA0302903), the National Natural Science Foundation of China (Grant Nos. 11674375, 11634015 and 11921004), the Strategic Priority Research Program of the Chinese Academy of Sciences (Grant No. XDB33010100), and the Postdoctoral Science Foundation of China.

\end{document}